\documentclass[11pt]{article}

\usepackage{putex}

\RequirePackage[a4paper,top=30.6mm,bottom=38.6mm,left=34.6mm,right=34.6mm,footskip=1.3cm]{geometry}
\usepackage{setspace}
\usepackage{booktabs}
\usepackage{cancel}
\usepackage{multirow}
\usepackage{comment}
\usepackage{bbold}
\usepackage{caption}
\usepackage{amsmath}
\usepackage{enumerate}
\usepackage{cite}
\usepackage{tensor}
\usepackage{slashed}
\usepackage[utf8]{inputenc}
\usepackage{rotating}
\usepackage{bigfoot}
\usepackage[
colorlinks=true,
linkcolor=black,
urlcolor=blue,
filecolor=black,
citecolor=red,
]{hyperref}

\usepackage[textsize=footnotesize]{todonotes}

\numberwithin{equation}{section}

\def \< {\left<}
\def \> {\right>}

\newcommand{\be}{\begin{equation}} \newcommand{\ee}{\end{equation}}
\newcommand{\bea}{\begin{eqnarray}}  \newcommand{\eea}{\end{eqnarray}}
\newcommand{\nn}{\nonumber}

\usepackage{amsfonts, amsthm}
\usepackage[english]{babel}
\usepackage{slashed}
\usepackage{mathrsfs}
\usepackage{amssymb}
\usepackage{color}

\newcommand{\ma}[1]{\mbox{$\mathcal{#1}$}}

\newcommand{\mrm}[1]{\mbox{$\mathrm{#1}$}}

\def\Im{{\mathrm{Im}}}
\def\Re{{\mathrm{Re}}}

\begin{document}

	\begin{center}        
		\Huge A positive metric over DGKT vacua 
	\end{center}
	
	\vspace{0.7cm}
	\begin{center}        
		{\large  Eran Palti\quad and\quad Nicol\`o Petri}
	\end{center}
	
	\vspace{0.15cm}
	\begin{center}  
		\emph{Department of Physics, Ben-Gurion University of the Negev,}\\
		 \emph{Be'er-Sheva 84105, Israel}\\[.3cm]
		\emph{}\\[.2cm]
	\end{center}
	
	\vspace{1cm}
	
	
	\begin{abstract}
	
	We study the notion of a metric over the space of AdS solution in string theory, leading to an associated distance between them. Such a distance is the idea underlying the AdS distance conjecture. We utilise the previously developed prescription for extracting such a metric: taking an off-shell quadratic variation of the string theory effective action and then evaluating it over the space of on-shell solutions. It was shown that this prescription leads to a well-defined positive metric over M-theory Freund-Rubin vacua. In this work, we use the same prescription to calculate the metric over type IIA DGKT vacua. These are much more involved, they have multiple flux parameters and exhibit scale separation. While it remains an open question whether these vacua exist as fully localised solutions of string theory, they are well-defined within the four-dimensional effective theory, which is all that is required for the calculation. We find that they also have a positive metric over them. Interestingly, this metric turns out to be independent of the many flux parameters in the solution, similarly to what happens for metrics over scalar field spaces. This non-trivial flux cancellation, as well as results from explicit vacua, lead us to propose a Swampland condition: that the metric over the space of vacua in quantum gravity, as defined by the above prescription, is always positive. 
	\end{abstract}
	
	\thispagestyle{empty}
	\clearpage
	
	\tableofcontents
	
	\setcounter{page}{1}

	\section{Introduction}
	\label{sec:intro}

There is a well-understood notion of a metric over the space of (supersymmetric) Minkowski vacua in string theory: the metric on the moduli space. This metric then defines a notion of distance between the vacua, and this notion of distance leads to certain constraints on the properties of theories at different distances. In particular, it has been conjectured that at infinite distances there is a tower of states which becomes light relative to the Planck scale \cite{Ooguri:2006in}. There is strong evidence for this proposal, see \cite{Palti:2019pca,vanBeest:2021lhn} for reviews, and \cite{Baume:2016psm,Klaewer:2016kiy,Blumenhagen:2018nts,Grimm:2018ohb,Grimm:2018cpv,Corvilain:2018lgw,Lee:2018spm,Lee:2019wij} for some of the initial work. 

In this paper we are interested in a metric, and a notion of distance, on non-Minkowski vacua, in particular, AdS vacua. The underlying idea behind such a notion of distance was proposed in \cite{Lust:2019zwm}. It was then significantly developed and made much more precise in \cite{Li:2023gtt}. In particular, a very specific prescription was introduced which leads to a metric over one-parameter families of AdS vacua in string theory. The prescription was to take a quadratic off-shell (so not constrained by the equations of motion) variation of the effective action, and then evaluate this on the family of on-shell solutions (to the equations of motion). More precisely, one promotes the parameters of the background, such as the radius of AdS or the flux numbers, to have a probe (infinitesimal) spatial dependence on the AdS radial direction $z$, and then extracts the quadratic variation of the action as the two-derivative terms acting on these parameters. The family of solutions to the equations of motion then gives a relation between the different parameters, say relating the AdS radius to the flux number (which strictly holds only for spatially constant values), that is then imposed on the off-shell spatial variations. After imposing such relations, the metric over that family of solutions is extracted as the coefficient multiplying the second derivative variations.

Let us note that there have been also different studies trying to understand AdS distance notions, related and unrelated to \cite{Lust:2019zwm}, in various settings. See \cite{Rudelius:2021oaz,Basile:2021mkd,Angius:2022aeq,Montero:2022ghl,Farakos:2023nms,Buratti:2020kda,Luben:2020wix,Li:2021utg,Collins:2022nux,Cribiori:2022trc,Shiu:2022oti,Cribiori:2023swd,Basile:2023rvm,Mohseni:2023ogd,Farakos:2023wps,Shiu:2023bay,Demulder:2023vlo} for an incomplete list.

In \cite{Li:2023gtt}, the metric was calculated over the space of Freund-Rubin solutions of M-theory. These are the families of AdS vacua of type AdS$_4\times S^7$ and AdS$_7 \times S^4$. Of particular interest was the sign of the metric. In the procedure above, the contribution to the metric coming from the variations of the AdS radius, the external conformal factor, led to a wrong-sign kinetic term, and so to a negative metric contribution. This is a manifestation of the famous conformal factor problem of Euclidean quantum gravity \cite{Gibbons:1976ue,Gibbons:1978ac,Marolf:2022ybi}. However, it was found that once the contribution to the metric coming from the on-shell variation of the internal sphere, and of the Freund-Rubin flux, was accounted for, the metric on the space of solutions became positive and so led to a well-defined distance. 

In this paper we calculate the metric on another, more involved, family of solutions in string theory: the DGKT vacua \cite{DeWolfe:2005uu}. These are vacua which are derived by studying the four-dimensional effective theory for massive type IIA supergravity on a Calabi-Yau orientifold. There exists an infinite family of such vacua, in which the internal volume, the string coupling, and the AdS radius are all controlled parametrically by the Ramond-Ramond four-form flux. In this sense they are similar to the Freund-Rubin vacua, and one can calculate the metric over this one-parameter family. But there are other features which make the vacua much richer than the Freund-Rubin solutions. For one, they involve multiple flux parameters. While only the four-form flux is parametrically adjustable, the other fluxes still can be chosen within a finite range. They also exhibit an unusual relation between the radius of the internal Calabi-Yau and the radius of AdS: they have parametric scale separation. Being string theory, rather than M-theory, vacua they also involve a scaling of the dilaton. All of these features will affect the calculation of the metric, and will therefore yield insights into the nature of such metrics. 

Unlike Freund-Rubin vacua, DGKT vacua have no known uplift to full ten-dimensional solutions. They can be uplifted to approximate solutions, where the orientifolds are treated in the smeared approximation \cite{Lust:2004ig,Acharya:2006ne}. By expanding about this smeared solution, it is possible to calculate a proper localised solution to first order in the expansion parameter, which is related to the (inverse of) the flux parameter \cite{Junghans:2020acz,Marchesano:2020qvg}. However, since this expansion is also the same expansion that is utilised for the supergravity approximation of string theory in the first place, it is not completely clear that such expansions are under control, and certainly the next order behaviour is difficult to capture. Overall, it is fair to say that whether the DGKT vacua are truly full solutions of string theory, which exhibit the same behaviour as the four-dimensional solutions, is an open question. In this paper, we will treat them as full solutions, defined by the four-dimensional effective action approach, and so by the smeared orientifold approximation.\footnote{Note that, as found in \cite{Marchesano:2019hfb,Marchesano:2020uqz}, there are a number of different vacua to the DGKT effective action. We will restrict to the original supersymmetric vacua found in \cite{DeWolfe:2005uu}.}

\subsection*{Summary of results and a Metric Positivity Conjecture} 

The calculation of the metric on DGKT vacua proceeds essentially the same as for the Freund-Rubin vacua in \cite{Li:2023gtt}. We extract the quadratic variations of the action with respect to the parameters in the family of solutions. There is a contribution from the conformal AdS factor variations and from the internal overall volume variations. Something new, relative to \cite{Li:2023gtt}, is we also have a contribution from dilaton variations. These three types of contributions are discussed in section \ref{metricvariations}. The flux variations are also similar to \cite{Li:2023gtt}, but with some additional subtleties because they are not spacetime-filling (not Freund-Rubin type). These subtleties are technical, rather than conceptual. We describe how to calculate the flux variation contributions to the metric in section \ref{sec:metoniia}.

The final answer we obtain for the metric is given in (\ref{DGKTmetric}). It is very similar to the Freund-Rubin metrics found in \cite{Li:2023gtt} in terms of its magnitude. Importantly, it is positive. This is so even though taking into account only the metric variations, the conformal AdS factor and the internal volume, it would have been negative. The additional dilaton and flux contributions flip the sign and render it positive, yielding a well-defined distance. 

The richness of the DGKT setting, relative to the Freund-Rubin one, also presents new insights into the metrics calculated this way. Apart from the parameter controlling the infinite family of solutions in DGKT, the four-form flux, there are multiple other flux parameters. For example, the NS three-form flux $q$ and the Romans mass $m$. We find that the metric over the space of solutions is independent of these other fluxes, they cancel out in the calculation. This is similar to what was found in \cite{Baume:2016psm}, where the metric measuring the distance along scalar field excursions with potentials was also found to be independent of the fluxes. This independence of any parameters is also a feature of moduli space metrics. It is intrinsically tied to the exponential behaviour of the tower of states in the distance conjecture appearing at sub-Planckian distances, as conjectured by the refined Distance Conjecture \cite{Baume:2016psm,Klaewer:2016kiy}. Indeed, it is precisely such flux cancellation which originally motivated the refined version of the conjecture. 

In the context of the metric on AdS space, the flux cancellation seems to yield support to the meaning of the metric, and in particular, to its required positivity. We propose that indeed the metric calculated this way has a physical meaning, and must therefore be positive, for any consistent family of vacua of string theory.

{\it Metric Positivity Conjecture} : The metric over solutions of quantum gravity, as calculated from an off-shell quadratic variation of the action restricted to the on-shell solutions, is always positive. 

In some sense, this was already tentatively proposed in \cite{Lust:2019zwm}. However, the work in \cite{Li:2023gtt} and in this paper has made it much more precise by giving a concrete prescription to calculate the metric. While the prescription has only been applied so far to relatively simple families of AdS vacua that are depending on just one parameter, the general procedure of quadratic variations in principle can be applied to all solutions. In practice, this is a difficult calculation, and we do not know how to implement it completely generally. But there are some relatively simple families of solutions in string theory, beyond Freund-Rubin and DGKT, for which our calculating procedure can be applied, and for these the conjecture implies very sharp predictions. We aim to perform these calculations for a spectrum of AdS families in future work.


The meaning of positivity of the metric for one-parameter families is clear, because the space is just a line. For multi-parameter families of solutions, it is not quite clear what it means. At the end, what we mean is that distances along paths in the parameter space are real, and therefore well-defined. We assume this needs to hold for any such path. But it could be that some restrictions on the path would need to be imposed in multi-parameter cases. Understanding the path dependence in multi-parameter families is a complicated and open question. 


\section{Metric and dilaton variations}
\label{metricvariations}

In this section we discuss metric and dilaton variations, in generic dimensions. The results of this section constitute an extension of Section 2 in \cite{Li:2023gtt}. In that context, a general formula for the metric over metric variations was obtained from external and internal volume variations. The vacua geometries considered were of the type AdS$_d\times Y_k$ with $Y_k$ a compact internal manifold. Here, we calculate how the metric is modified through the presence of dilaton variations. 

The main result of this section is a general formula for such a metric, which includes dilaton variations and describes their mixing with variations of the internal volume. This formula will be utilised in the special setting of DGKT vacua to calculate the metric over them.

\subsection{External volume variations}
\label{externalvariations}

The metric over metric variations involves the notion of external volume variation from Weyl rescalings \cite{Lust:2019zwm,Li:2023gtt}. The idea is starting from AdS space\footnote{We point out that this analysis holds also for de Sitter space.} in generic $d$ dimensions and deforming the geometry as follows
\begin{equation}\label{external}
 ds_d^2=e^{2\sigma}\,d\hat s^2_d\,,
\end{equation}
where we take $d\hat s^2_d$ to be a $d$-dimensional metric on unitary AdS and $\sigma$ is a function over AdS. When $\sigma$ is a constant, this variation connects two AdS spaces of different vacuum energy $\Lambda$, with $\Lambda=e^{-2\sigma}\,\hat\Lambda$. One can then use \eqref{external} to introduce a notion of metric over the space of the vacuum energy variations. This is done by promoting $\sigma$ to have some spacetime dependence, and then extracting a kinetic term from the Einstein-Hilbert action. Such a procedure implies that the equations of motion of the theory are no longer satisfied. In other words, after promoting a global parameter to a local one, the AdS geometry gets (infinitesimally) deformed. Following \cite{Li:2023gtt}, we will refer to this procedure as going {\it off-shell} with respect to the vacuum geometry by means a {\it gauging} of $\sigma$. Once the metric is extracted, we can go back {\it on-shell} taking constant variations.

The off-shell action for deformations \eqref{external} has been worked out in \cite{Li:2023gtt}. The derivation uses the standard formula of the Ricci scalar under Weyl rescalings,
\begin{equation}\label{Riccitransformed}
\begin{split}
 &R_d=e^{-2\,\sigma}\,\left(\hat R_d-(d-1)(d-2)\hat g_d^{mn} \partial_m \,\sigma \partial_n \,\sigma-2(d-1)\,\hat \nabla_d^2\,\sigma   \right)\,,\\
 & \hat \nabla_d^2\,\sigma=\frac{1}{\sqrt{-\hat g_d}}\,\partial_m\,\left(\sqrt{-\hat g_d}\,\hat g_d^{mn}\,\partial_n\,\sigma    \right)\,.
 \end{split}
\end{equation}
We can obtain the off-shell action by using \eqref{Riccitransformed} in the Einstein-Hilbert action\footnote{We work in Planck units $M_p=1$, with $M_p$ Planck mass in lower $d$ dimensions.} $S_d =\frac12\int d^dx\sqrt{-g_d}\,R_d$ and performing a partial integration of the laplacian. This leads to the following expression of the Einstein-Hilbert term \cite{Li:2023gtt}
\begin{equation}
\begin{split}\label{EHoffshell}
 &S_d=\frac12\int d^dx\sqrt{-g_d}\,\left(\tilde R_d-K_{\sigma\sigma}(\partial \,\sigma)^2 \right)\qquad \text{with}\qquad K_{\sigma\sigma}=-(d-1)(d-2)\,,
  \end{split}
\end{equation}
where we omitted total derivatives and we introduced the notation $\tilde R_d=e^{-2\sigma}\hat R_d$. It is important to point out that the above action is defined over off-shell configurations. In other words the action \eqref{EHoffshell} has been written absorbing\footnote{In \eqref{EHoffshell} we use the notation $(\partial \sigma)^2=g_d^{mn}\partial_m\sigma\partial_n\sigma$.} the factor $e^{d\sigma}$ within the volume $\sqrt{-g_d}$ and using the relation $e^{-2\sigma}\,\hat g_d^{mn}=g_d^{mn}$ in the kinetic term for $\sigma$.

The coefficient $K_{\sigma\sigma}$ is the metric element over external volume variations. We point out that this factor is negative definite. This is related to the negative contribution to the gravity action from the conformal mode of the metric, which is at the origin of the conformal factor problem \cite{Gibbons:1976ue,Marolf:2022ybi}. 

\subsection{Internal volume variations and the dilaton}
\label{internalvar}

We now turn to internal volume variations. To this aim we consider vacua of the type AdS$_d\times Y_k$ with $Y_k$ a compact manifold. We will include also the coupling to a dilaton $\Phi$. Given these variations, in this section we extract their contributions to the metric over vacua variations. We will keep constant, for the moment, the external volume of AdS. In the next section we will the combine external and internal variations to get the general metric formula.

We can consider the following metric deformation,
\begin{equation}\label{internal}
  ds_{d+k}^2=d s^2_d+e^{2\tau}\,d\hat s^2_k\,,
\end{equation}
where $\tau$ is a function over AdS$_d$ and $d\hat s^2_k$ is the metric over $Y_k$. We want now to compute the kinetic terms associated to $\tau$ and $\Phi$, to this aim we proceed by reducing the $D$-dimensional action to $d$-dimensions. We can then start with the standard {\itshape string frame} action given by
\begin{equation}
 \begin{split}\label{EHdilatonAction}
  S=\frac{1}{2\kappa_{d+k}^2}\int d^{d+k}x\sqrt{-g}\,e^{-2\Phi}\left(R+ 4g_d^{mn} \partial_m \Phi\, \partial_n\Phi \right)\,,
 \end{split}
\end{equation}
where $\kappa^2_{d+k}$ is the gravitational coupling in $d+k$ dimensions.
We can expand the Ricci scalar using the standard dimensional reduction prescription\footnote{For the derivation of equation \eqref{reductionR} see Appendix A in \cite{Li:2023gtt}.}
\begin{equation}\label{reductionR}
 R= R_d+e^{-2\tau}\, R_k-k(k+1)\,  g_d^{mn} \partial_m \,\tau \partial_n\tau-2k\, \nabla_d^2\,\tau\,,
\end{equation}
with $ \nabla_d^2$ given in \eqref{Riccitransformed}. We point out that the contractions in the above equations are expressed only in terms of $d$-dimensional indices $(m,n, \dots)$ since $\tau$ is varied over AdS$_d$.
We proceed by integrating by parts the Laplacian $\nabla_d^2\tau$, including the volume and dilaton factors in the derivative. The result is the following
\begin{eqnarray}
  \label{internaleh}
  S&=&\frac{1}{2\kappa_{d+k}^2}\int d^{{d+k}}x\sqrt{- g_d}\sqrt{\hat g_k}\,e^{k\,\tau-2\Phi}\,\bigl( R_d+e^{-2\tau}\,\hat R_k+4g_d^{mn} \partial_m \Phi\, \partial_n\Phi\\
  &-&k(k+1)\,g_d^{mn} \partial_m \,\tau \partial_n\tau \nn 
  +2k\,e^{-k\tau+2\Phi}\,\partial_m\bigl(e^{k\,\tau-2\Phi}\bigr)\, g_d^{mn}\partial_n\tau\bigr) \nn\\
  &=&\frac{1}{2\kappa_{d+k}^2}\int d^{d+k}x\sqrt{- g_d}\,\sqrt{\hat g_k}\,e^{k\,\tau-2\Phi}\,\bigl( R_d+e^{-2\tau}\,\hat R_k+k(k-1)\, g_d^{mn} \partial_m \,\tau \partial_n\tau \nn \\
 &-&4k\, g_{d}^{mn}\partial_m \Phi\,\partial_n\tau+4g_d^{mn} \partial_m \Phi\, \partial_n\Phi\bigr)
\end{eqnarray}

We now need to go to the lower-dimensional theory in the Einstein frame. To do so, let us first perform a Weyl rescaling to a $d$-dimensional metric $d s_{E}^2$ as  
\begin{equation}
 \begin{split}\label{WeylRescaling}
  &d s_{d}^2= e^{2D}\,ds_{E}^2\,,\\
  & R_d= e^{-2D}\,\left(R_{E}-(d-1)(d-2)g_{E}^{mn} \partial_m D \,\partial_nD-2(d-1)\,\nabla_E^2\,D   \right)\,,
 \end{split}
\end{equation}
where the laplacian is expressed in terms of $d s^2_{E}$. We can thus rewrite the Einstein-Hilbert action in the following form
\begin{equation}
 \begin{split}
  S&=\frac{1}{2\kappa_{d+k}^2}\int d^{d+k}x\sqrt{- g_{E}}\sqrt{\hat g_k}e^{(d-2)D}\,e^{k\tau-2\Phi}\bigl( R_{E}+e^{-2\tau+2D}\hat R_k+k(k-1) g_{E}^{mn} \partial_m \tau \partial_n\tau \\
  &-4k\, g_{E}^{mn}\partial_m \Phi\,\partial_n\tau+4g_E^{mn} \partial_m \Phi\, \partial_n\Phi-(d-1)(d-2) g_{E}^{mn} \,\partial_m D \partial_nD-2(d-1)\,\nabla_E^2\,D \bigr)\,.
 \end{split}
\end{equation}
The $d$-dimensional Einstein frame is the frame where field variations are chosen as
\begin{equation}\label{ConditionEF}
 D=\frac{2\Phi-k\tau}{d-2}\,.
\end{equation}
Imposing \eqref{ConditionEF}, the laplacian $\nabla_E^2\,D $ becomes a total derivative, leading to the following $d$-dimensional action\footnote{We use the notation $(\partial \,\tau)^2= g_{E}^{mn}\partial_m \tau \partial_n\tau$.},
\begin{equation}\label{EinsteinFrameActioninternal}
 \begin{split}
  &S_d=\frac{1}{2}\int d^dx\sqrt{- g_{E}}\left(R_{E}+e^{-2\tau+2D}\,\hat R_k- K_{\tau\tau}\,( \partial\tau)^2-2K_{\Phi\,\tau}(\partial \Phi)(\partial\tau) -K_{\Phi\Phi}( \partial \Phi)^2\right),
 \end{split}
\end{equation}
where we fixed $\kappa_{d+k}^{2}$ such that the $d$-dimensional action is in Planck units. In the above expression we have introduced the coefficients
\begin{equation}
\begin{split}\label{internalMetric}
 & K_{\tau\tau}=k^2\,\left(\frac{d-1}{d-2}\,-\frac{k-1}{k}\right),\qquad K_{\Phi\Phi}=\frac{4}{d-2}\,,\qquad K_{\Phi\,\tau}=-\frac{2k}{d-2}\,,
\end{split}
\end{equation}
which are the metric elements over internal volume and dilaton variations. Finally we observe that, employing the field $D$ introduced in \eqref{ConditionEF}, we can cast the above action in a diagonal form
\begin{equation}\label{EinsteinFrameActioninternal4D}
 \begin{split}
  &S_d=\frac{1}{2}\int d^dx\,\,\sqrt{- g_{E}}\,\left(R_{E}+e^{-2\tau+2D}\,\hat R_k-K_{DD}\,( \partial D)^2-\tilde K_{\tau\tau}\,( \partial \,\tau)^2\right)\,,
 \end{split}
\end{equation}
where the metric coefficients in the new basis are given by
\begin{equation}\label{KDD}
 K_{DD}=(d-2)\,,\qquad \qquad \tilde K_{\tau\tau}=k\,.
\end{equation}
The modulus $D$ is usually called the $d$-dimensional dilaton. In what follows we will mainly use the action \eqref{EinsteinFrameActioninternal}, expressed in term of the 10d dilaton $\Phi$. 

\subsection{Combined variations and $\mrm{AdS}_4\times Y_6$ vacua}
\label{generalvar}

Let us now consider the combination of external and internal variations. When the variations are constant, it is clear that they can be taken independently. As is explicitly showed in \cite{Li:2023gtt}, this fact remains true also when the variations are gauged. The total action is just the sum of the two contributions. In this present case we have also dilaton variations, but the situation is unchanged.
We can thus consider the following metric variations in the {\itshape string frame}
\begin{equation}
  ds_{d+k}^2=e^{2\sigma}d\hat s^2_d+e^{2\tau}\,d\hat s^2_k\,,
\end{equation}
where $d\hat s^2_d$ and $d\hat s^2_k$ are the metrics of unitary AdS$_d$ and $Y_k$. The lower-dimensional action resulting from general metric variations has the following form
\begin{equation}\label{genvaraction4D}
 \begin{split}
  S_d=\frac{1}{2}\int d^dx\,\,\sqrt{- g_{E}}\,&\bigl(\tilde R_{E}+e^{-2\tau+2D}\,\hat R_k-K_{\sigma\sigma}\,( \partial \,\sigma)^2\\
  &- K_{\tau\tau}\,( \partial\tau)^2-2K_{\Phi\,\tau}(\partial \Phi)(\partial\tau) -K_{\Phi\Phi}( \partial \Phi)^2\bigr),
 \end{split}
\end{equation}
where we introduced the notation $\tilde R_{E}=e^{-2\sigma}\hat R_{E}$. The metric elements are given by
\begin{equation}
\begin{split}
 &K_{\sigma\sigma}=-(d-1)(d-2)\,,\qquad K_{\tau\tau}=k^2\,\left(\frac{d-1}{d-2}\,-\frac{k-1}{k}\right)\,,\\ &K_{\Phi\Phi}=\frac{4}{d-2}\,,\qquad \qquad \qquad K_{\Phi\,\tau}=-\frac{2k}{d-2}\,.
 \end{split}
\end{equation}

A central feature of the procedure presented in \cite{Li:2023gtt} is reducing the off-shell action by using consistency conditions among variations coming from the equations of motion. We call these constraints {\itshape on-shell conditions}. This idea takes inspiration from string theory vacua, where the equations of motion relate the variations within a family of AdS solutions. In order to reduce the off-shell action one needs to assume that these relations holds also for  gauged variations. The simplest example of such procedure is proposed in \cite{Li:2023gtt} with Freund-Rubin vacua.

Following this strategy, one can obtain the metric over metric and dilaton variations, at least for the case of one-parameter families of AdS solutions, in general. Let us suppose that the equations of motion are satisfied by the following scalings
\begin{equation}\label{generalscalings}
 e^{2\sigma}\sim n^a\,,\qquad e^{2\tau}\sim n^b \qquad e^{\Phi}\sim n^c,
\end{equation}
where $n$ is a constant parameter featuring the solution.
When the above variations are gauged, $n$ is a function over AdS and from \eqref{genvaraction4D} we can write the action
\begin{equation}\label{genvaractionN}
 \begin{split}
  &S_d=\frac{1}{2}\int d^dx\,\,\sqrt{- g_{E}}\,\left(\tilde R_{E}+n^{\frac{4c-b(d+k-2)}{d-2}}\,\hat R_k-K^{\text{metric}}_{nn}\,( \partial \log n)^2\right),
 \end{split}
\end{equation}
where we introduced the total metric over the variations,
\begin{equation}\label{generalmetricabc}
  K^{\text{metric}}_{nn}=-\frac{a^2(d-2)^2(d-1)+8bck-b^2k(d+k-2)-16c^2}{4(d-2)}\,.
\end{equation}

The case of Freund-Rubin vacua is realized by setting $a=b$ and $c=0$. From the formula above it is easy to recover the values for AdS$_5\times S^5$, AdS$_4\times S^7$, and AdS$_7\times S^4$ written in eq. (2.22) in \cite{Li:2023gtt}. Using the scalings \eqref{generalmetricabc}, the metric on metric variations for the three Freund-Rubin vacua can be written\footnote{Using $n=e^{\frac2a\,\sigma}$, the metric elements take the form of (2.22) in \cite{Li:2023gtt}, namely $K^{\text{metric}}_{\sigma\sigma}= \frac{4}{3},\,\frac{51}{2}\,,-\frac{114}{5}$.} as $K^{\text{metric}}_{nn}= \frac{a^2}{3},\,\frac{51a^2}{8}\,,-\frac{57a^2}{10}$.

Let's now focus on type II vacua of the type AdS$_4\times Y_6$. In this case metric variations in the string frame have the following form
\begin{equation}
 ds^2_{10}=e^{2\sigma}\,d\hat s_4^2+e^{2\tau}\,d\hat s^2_6\,,
\end{equation}
where $d\hat s_4^2$ is the metric of unit AdS$_4$ and $d\hat s^2_6$ is the metric over $Y_6$. We can restrict the total metric over metric variations \eqref{generalmetricabc} to the case of $d=4$, $k=6$. This leads to the formula
\begin{equation}\label{4Ddistance}
 K^{\text{metric}}_{nn}=-\frac12\,\left(3a^2-12b^2+12bc-4c^2   \right)\,.
\end{equation}

In this work we are interested in DGKT vacua \cite{DeWolfe:2005uu}. These arise from a class of compactifications of massive type IIA over Calabi-Yau orientifold. These vacua constitute a one-parameter family of solutions with the following scalings:
\begin{equation}\label{DGKTscalings}
 e^{2\sigma}\sim n^{3/2}\,,\qquad e^{2\tau} \sim n^{1/2} \qquad e^{\Phi}\sim n^{-3/4}\,,
\end{equation}
where $n$ describes the scaling of the internal RR 4-flux in the limite of large internal volume.
Evaluating formula \eqref{4Ddistance} over the above scalings, we obtain the result
\begin{equation}
 K^{\text{metric}}_{nn}=\frac32\,,
\end{equation}
which give a positive-definite metric over metric and dilaton variations of DGKT vacua. We point out that this result is only partial contribution to the full metric, which includes a contribution from flux variations. We study this contribution in the next section.

\section{The metric on type IIA orientifold vacua}
\label{sec:metoniia}

Type IIA orientifold compactifications define a class of four-dimensional $\ma N=1$ effective theories\footnote{We refer to \cite{DeWolfe:2005uu} for an account within the context of this work, and to \cite{Grimm:2004ua} for the general original work. A short review of main properties of moduli spaces of type IIA orientifold compactifications is presented in Appendix \ref{appendixModuliSpace}.}. 
Of these, DGKT vacua \cite{DeWolfe:2005uu} are a specific family of AdS$_4$ solutions. There are two types of fluxes in these vacua. There are fluxes which are bounded by the tadpole constraints, such as the Romans mass $m$ and internal NS-NS flux $H_3$. There is also the RR flux $F_4$, which is unrestricted by tadpoles, and so can be varied parametrically. This flux defines for us the one-parameter family of DGKT vacua.
 

A similar family of solutions parameterised by a single flux parameter are Freund-Rubin vacua. In this case, it was shown in \cite{Li:2023gtt} that flux variations give a non-trivial contribution to the metric over vacua. 
In this section, we calculate, in a similar way, the contribution to the metric from variations of the $F_4$ internal flux in DGKT vacua. This is the unique flux we will consider since all the others are fixed to restricted set of values by the tadpole conditions. This will lead us to the total metric over the field variations. As was the case for Freund-Rubin vacua, the procedure to compute such a metric is very subtle since metric and flux variations combine non-trivially, in a way that the total metric is not the simple sum of their contributions. 

\subsection{Calabi-Yau metric variations}

 Let us consider the metric variations of Calabi-Yau orientifolds. These have 3-cycles and 2-cycles, which are dual to 4-cycles. We first point out that the presence of the orientifold strongly restricts the set of possible variations of the internal geometry. A more detailed analysis on this topic can be found in Appendix \ref{appendixModuliSpace}.
 
We consider first variations associated to the K\"ahler moduli space. These variations are related to the 2-cycles and they are typically obtained by retaining only those moduli defined by odd (1,1)-forms under the orientifold (for more details see Appendix \ref{KahlerModuli}). Specifically, we can introduce $h^{1,1}_-$ complex moduli $t^a=b^a+i v^a$ defined by the following expansions
\begin{equation}
 J=v^aw_a\qquad \text{and} \qquad B_2=b^aw_a \,,
\end{equation}
where $\{w_a\}$ is a basis of odd harmonic (1,1)-forms with $a=1,\dots,h^{1,1}_-$. The $b^a$ fields are axions associated to the $B_2$ field.
The moduli $v^a$ describe the volume variations of 2-cylces and they can be used to define the total volume of the Calabi-Yau as
\begin{equation}
 \text{vol}=\frac16\,\kappa_{abc}v^av^bv^c \qquad \text{with}\qquad \kappa_{abc}=\int w_a\wedge w_b\wedge w_c\,,
\end{equation}
where $\kappa_{abc}$ is the intersection number of the Calabi-Yau.

Let us now consider, for simplicity, varying all the 2-cycles by the same volume variation. We can describe such variations by introducing an overall modulus $\tau$ as 
\begin{equation}\label{taudef}
 v^a=e^{2\tau}\,\hat v^a \qquad \text{with} \qquad \text{vol}=e^{6\tau}\,,
\end{equation}
where $\hat v^a$ are constant parameters defining the unitary volume $\hat{\text{vol}}\equiv \frac16\kappa_{abc}\hat v^a \hat v^b \hat v^c=1$. Under this assumption, we can factorize volume variations out from the metric
\begin{equation}\label{internalCYvariations}
 ds^2_6=e^{2\tau}\,d\hat s^2_6\,,
\end{equation}
where $d\hat s^2_6$ is the metric over the Calabi-Yau with unitary volume. We thus obtain the same type of internal volume variations discussed in Section \ref{internalvar}.

The variations of 3-cycles that survive under the orientifold are described by $h^{2,1}+1$ complex moduli. 
More precisely, $h^{2,1}$ complex moduli can be defined expanding the fundamental 3-form $\Omega$ and the gauge potential $C_3$ as \cite{Grimm:2004ua}
\begin{equation}\label{omegaexpansion}
 \Omega=(\Re Z^{k}\alpha_{k}+i\Im Z^\lambda\alpha_{\lambda})  - (\Re \ma F_{\lambda}\beta^{\lambda}+i\Im \ma F_k\beta^{k})\qquad \text{and} \qquad C_3=\xi^k\alpha_k-\tilde \xi_\lambda \beta^\lambda\,,
\end{equation}
where $\{\alpha_k, \beta^\lambda \}$ and $\{\alpha_\lambda, \beta^k \}$ are two bases of harmonic 3-forms, respectively even and odd under the orientifold. The indices $(k, \lambda)$ vary in the intervals $k=0,1,\dots,h$ and $\lambda=h,\dots, h^{2,1}$, where $h$ is a basis-dependent value\footnote{See Appendix \ref{hyperModuli} for more details on this point.} fixed by the number of even $\alpha$ forms. The moduli defined by $\Omega$ variations come from splitting the homogenous coordinates as $(Z^{\hat K}, \ma F_{\hat L})=( Z^{k}, Z^{\lambda} , \ma F_k, \ma F_\lambda)$. The orientifold action projects out half of these fields through the relations $\Im\, Z^k=\Re \,\ma F_k=\Re \,Z_\lambda=\Im \ma \,F_\lambda=0$. If we now consider the axions, half of them must be excluded since $C_3$ is even under the orientifold. In \eqref{omegaexpansion} we called the surviving axions $\xi^k$ and $\tilde \xi_\lambda$. Finally, we point out that also the universal hypermultiplet is cut in half. The moduli retained after the orientifold are the 10d dilaton $\Phi$ and one axion $\xi$. 

We may conclude this section by writing the total K\"ahler potential associated to the $\ma N=1$ moduli space. As it is shown in \cite{DeWolfe:2005uu, Grimm:2004ua}, this can be written in the very simple form
\begin{equation}\label{totalKahler}
 K=K^K+K^Q\qquad \text{with}\qquad e^{K^K}=\frac18\,e^{-6\tau}\,,\qquad e^{K^Q}=e^{4D}\,,
\end{equation}
where $e^D=e^{\Phi-3\tau}$ is the 4d dilaton, introduced in \eqref{ConditionEF}. In the above expression we used the K\"ahler potentials $K^K$ and $K^Q$ given in equations \eqref{KahlerKK} and \eqref{KahlerKQ}.

\subsection{Fluxes and on-shell conditions}

Let us discuss the on-shell conditions for DGKT vacua. As we said, these are the set of algebraic equations that stabilize field variations at their vacuum expectation values. In other words, we look at the conditions under which the potential is extremized. We point out that these conditions are not unique, particularly they can preserve supersymmetry or not \cite{Narayan_2010}. We will consider the standard $\ma N=1$ vacua of \cite{DeWolfe:2005uu}.

We consider the following type IIA fluxes \cite{DeWolfe:2005uu},
\begin{equation}\label{onshellfluxes}
\begin{split}
 & H_3=q^\lambda \alpha_\lambda-p_k\beta^k\,,\qquad F_0=m\,,\qquad F_4=e_a\tilde w^a\,.
 \end{split}
\end{equation}
The above fluxes are decomposed along $\{\alpha_\lambda, \beta^k\}$ and $\{\tilde w^a\}$, the bases of odd harmonic 3-forms and even (2,2)-forms. The latter can be defined as the Hodge dual to the basis $\{w_a\}$ through the identities \eqref{dualomega}. 

We can start by studying the stabilization of the K\"ahler moduli $(b^a, v^a)$. The flux configuration written above requires that the axions $b^a$ vanish\footnote{For a compactification with non-trivial $b^a$, one would need internal 2-fluxes $F_2=m^a w_a$. For simplicity we will consider compactifications with $m^a=0$ and $b^a=0$.}.
Let us then consider volume variations of 2-cycles $v^a$. As shown in \cite{DeWolfe:2005uu}, these fields are entirely stabilized by fluxes $e_a$. Specifically, at the vacuum these variations satisfy the following algebraic equation \cite{DeWolfe:2005uu}
\begin{equation}\label{onshellCondFlux}
 \text{on-shell:} \qquad \qquad e_a=-\frac{3m}{10}\,\kappa_{abc}v^bv^c\,.
\end{equation}
This relation leads to the expression for the superpotential at the vacuum $W=\frac{2im}{15}\,\kappa_{abc}v^av^bv^c$ given in \cite{DeWolfe:2005uu}. The vacuum energy can be obtained using the standard formula of $\ma N=1$ supergravity,
\begin{equation}\label{vacenergy}
 \Lambda=-3e^{K}|W|^2=-\frac{6m^2}{25}\,e^{4\Phi-6\tau}\,,
\end{equation}
where we used equation \eqref{taudef} to rewrite $\kappa_{abc}v^av^bv^c=6\,e^{6\tau}$ and the total K\"ahler potential $K$ is given in \eqref{totalKahler}. 

The expression for the vacuum energy \eqref{vacenergy} can be used to relate the internal volume variations to the external ones. From Section \ref{generalvar}, we know that the variations of the AdS$_4$ volume $ds^2_4=e^{2\sigma}d\hat s^2_4$, defined in the 10d string frame, can be rewritten in the 4d Einstein frame as
\begin{equation}\label{extvariationsDGKT}
 ds^2_E=e^{-2D}\,ds^2_{4}=e^{2\sigma-2D}\,d\hat s^2_4\qquad \text{with}\qquad e^{2D}=e^{2\Phi-6\tau}\,,
\end{equation}
where $d\hat s^2_4$ is the metric of unitary AdS$_4$ and $ds^2_E$ is the 4d Einstein frame metric. By definition, the vacuum energy in a $d$-dimensional gravity theory is given by $\Lambda=-\frac{1}{2}(d-1)(d-2)R_{\text{AdS}}^{-2}$. For metrics of the type \eqref{extvariationsDGKT}, this relation takes the form $\Lambda=-3e^{-2\sigma+2D}$. If we now compare this result with \eqref{vacenergy}, we find the following fundamental relation 
\begin{equation}\label{onshell:dilaton}
 \text{on-shell:}\qquad\qquad \Phi=-\sigma+\frac12\log\left(\frac{25}{2m^2}\right)\,.
\end{equation}

Let us consider now the remaining conditions coming from the complex structure moduli space. The dilaton variations are fixed at the vacuum in terms of internal variations by the condition $ e^{-\Phi}=\frac{8m}{5 q}\,e^{3\tau}$ \cite{DeWolfe:2005uu}. This lead to the following relation between external and internal variations
\begin{equation}\label{onshell:tau}
 \text{on-shell:} \qquad \qquad \tau=\frac{\sigma}{3}+\frac16\log\left(\frac{q^2}{32}\right)\,.
\end{equation}
We notice that in the above relation the scale separation between the AdS$_4$ vacuum and the Calabi-Yau is now fully explict (in a vacuum geometry without scale separation we would have $\tau=\sigma$).

The constant parameter $q$ is defined in terms of the non-zero complex structure moduli introduced in \eqref{omegaexpansion} and the $H_3$ fluxes $(q^\lambda, p_k)$ as\footnote{When any $q^\lambda$ or $p_k$ vanishes, the corresponding field $\Im Z^{\lambda}$ or $\Im \ma F_{k}=0$ vanishes \cite{DeWolfe:2005uu}.} \cite{DeWolfe:2005uu}
\begin{equation}\label{qdef}
 \text{on-shell:} \qquad q\equiv e^{-K^{\text{cs}}/2}\frac{p_{k_1}}{\Im \ma F_{k_1}}=e^{-K^{\text{cs}}/2}\frac{p_{k_2}}{\Im \ma F_{k_2}}=\cdots=e^{-K^{\text{cs}}/2}\frac{q^{\lambda_1}}{\Im  Z^{\lambda_1}}=\cdots\,,
\end{equation}
where $K^{\text{cs}}$ is the K\"ahler potential on the complex structure moduli space restricted to moduli surviving under the orientifold. Its explicit form is given in \eqref{complexStructureNT}.

Let's finally consider the axions $\xi^k$, $\tilde\xi_\lambda$. These appear in the unique relation of the form 	\cite{DeWolfe:2005uu}
\begin{equation}\label{axionxxi}
 \text{on-shell:}\qquad \qquad p_k \xi^k-q^\lambda \tilde \xi_\lambda=e_0\,,
\end{equation}
where is a flux parameter associated to the dual electric 6-flux. Since this a single relation for $h^{2,1}+1$ axions, it follows that only a single combination of the axions $\xi^k$, $\tilde\xi_\lambda$ gets stabilized by fluxes, while the remaining ones are not. As explained in \cite{DeWolfe:2005uu}, this issue can be potentially resolved by D2 brane instantons.

Even if in the general case we cannot stabilize all the axions, one can focus on concrete compactifications where this issue is not present. This is the case of the orbifold $T^6/\mathbb{Z}^2_3$, which is featured by $h^{2,1}=0$. In this compactification the complex structure moduli space is trivial and there is a single axion $\xi$ from the universal hypermultiplet. Let's review the main features of this model.

\subsubsection{The case of the orbifold $T^6/\mathbb{Z}^2_3$}

The orientifold compactification over the orbifold $T^6/\mathbb{Z}^2_3$ was studied in \cite{DeWolfe:2005uu}. In this model, the invariance under the action of the two orbifold symmetries $\mathbb{Z}_3\times\mathbb{Z}_3$ implies that all the off-diagonal terms in the internal metric vanish, leading to the following metric for internal variations
\begin{equation}
\begin{split}
 &ds_6^2=2(\sqrt 3\,\kappa)^{1/3}\left(v^1 ds^2_{T^2_1}+v^2 ds^2_{T^2_1}+v^3ds^2_{T^2_3}\right)=2(\sqrt 3\,\kappa)^{1/3}e^{2\tau}\,d\hat s^2_6\,,\\
 \end{split}
\end{equation}
where $d\hat s^2_6$ is the metric over $T^6/\mathbb{Z}^2_3$ with unitary volume\footnote{The metric is normalized following the conventions of \cite{DeWolfe:2005uu}.} and $\kappa\equiv \kappa_{123}$ is the intersection number. The three moduli $v^a$ describe the size of the three tori $T^2_a$. All the other variations of the internal metric are turned off. Moreover, since the 10d action is quadratic in the $b^a$ axions from the $B_2$ field, it follows that they also can be taken vanishing. We can introduce a basis of harmonic forms as
\begin{equation}
 w_a=(\sqrt 3\,\kappa )^{1/3}\,\text{vol}_{T^2_a}\qquad \text{and} \qquad \kappa=\int w_1\wedge w_2\wedge w_3\,,
\end{equation}
where $\text{vol}_{T^2_a}=i dz^a\wedge d\bar z^a$ with $\{z^a\}$ complex coordinates over the $T^2_a$.
The dual basis $\tilde w^i$ can be obtained from $w_i$ by applying \eqref{dualomega}.

In this model $h^{2,1}=0$, namely the only non-vanishing complex structure moduli are  the dilaton $\Phi$ and its axion $\xi$. The fundamental 3-form and the 3-form gauge potential \eqref{omegaexpansion} can be decomposed as
\begin{equation}
 \Omega=\frac{1}{\sqrt{2}}(\alpha_0+i\beta^0)\,,\qquad C_3=\xi\alpha_0\,,
\end{equation}
with $\int \alpha_0\wedge \beta^0=1$.
The non-trivial fluxes in \eqref{onshellfluxes} take the form
\begin{equation}
\begin{split}\label{fluxesZ3}
 &H_3=-p\, \beta^0\,,\qquad F_0=m\,,\qquad F_4=e_i\tilde w^i\,.
 \end{split}
\end{equation}
With this flux configuration, one can see that this model does not suffer from the issue of the stabilization of the complex structure axions mentioned above. Specifically, here we just have the axion $\xi$ belonging to the hypermultiplet. The on-shell condition \eqref{axionxxi} boils down to  $\xi=\frac{e_0}{p}$. 

The flux $p$ in \eqref{fluxesZ3} is directly related to the parameter $q$ defined in \eqref{qdef}. In fact, using that in this model $\Im\, \ma F_0=-\frac{1}{\sqrt 2}$ and $K^{\text{cs}}=0$, one has $q=-\sqrt 2 \,p$.
The fluxes $H_3$ and $F_0$ are fixed by a single tadpole condition related to the orientifold O6 plane. This takes a particularly explicit form in the $T^6/\mathbb{Z}^2_3$ model\footnote{In general compactifications one has $h^{2,1}+1$ tadpole conditions.} given by \cite{DeWolfe:2005uu},
\begin{equation}\label{tadpole}
 m\,p=-2^{3/4}\pi^{1/8}\,\kappa_{10}^{1/4}\,,
\end{equation}
where $\kappa_{10}^2$ is the 10d gravitational coupling. The above relation must be compatible with the quantization conditions for $F_0$ and $H_3$ fluxes, implying that it can be satisfied by very specific values of quantized fluxes \cite{DeWolfe:2005uu}. From this, it follows that the unique fluxes free to vary are the parameters $e_i$, associated to $F_4$.

Finally, we can write the scalar potential of the 4d effective theory \cite{DeWolfe:2005uu}
\begin{equation}\label{T6potential}
 V=\frac{m^2}{2}\,e^{4\Phi-6\tau}+\frac{1}{2}e_a^2v_a^2\,e^{4\Phi-18\tau}+\frac{p^2}{4}e^{2\Phi-12\tau}-\sqrt2\,|mp|e^{3\Phi-9\tau}\,,
\end{equation}
where the first three terms respectively belong to $F_0$, $F_4$, and $H_3$ fluxes, while the last one is obtained by reducing the orientifold action.

\subsection{Flux variations}
\label{sec:fluxvariations}

We are now ready to study flux variations in DGKT vacua. As mentioned above, DGKT vacua constitute an infinite family of AdS$_4$ solutions parameterized by the internal flux $F_4=e_i\tilde w^i$. We will consider variations of this flux only, giving a one-parameter family.
We apply the same logic we adopted for metric variations in Section \ref{metricvariations}, that is, taking off-shell variations of the physical parameters of the vacuum solution. For the $F_4$ flux, this is amounts to taking continuous variations over the flux parameters $e_i$. Of course, in the on-shell full string vacuum, the fluxes are quantised. We can think of the string vacua as lying at discrete points along a continuous line, and we are calculating the metric on the line.

The $F_4$ flux in DGKT vacua is entirely threading the dual 4-cycles within the Calabi-Yau, namely it is purely magnetic. In this case, the prescription for flux variations is formulated from the electric dual flux \cite{Li:2023gtt}.
We can define the flux variations from a gauge potential $C_5$, associated to the electric dual 6-flux $F_6$. The variations of the $F_4$ flux are then obtained by dualizing the $F_6$.

Let us start our analysis by looking with more detail at the on-shell configurations of the $F_4$ flux. We can introduce an (undeformed) gauge potential $\hat C_5$ and its corresponding flux as
\begin{equation}
 \hat F_6=d\hat C_5\qquad \text{with}\qquad \hat F_4= \star_{10} \hat F_{6}\,.
\end{equation}
In order to work explicitly as possible we can introduce the Poincar\'e patch on AdS$_4$,
\begin{equation}
\begin{split}\label{AdS4cord}
 &d\hat s^2_{4}=z^{-2}\bigl(ds^2_{M_3}+dz^2\bigr),\\
 &\text{vol}_{4}=z^{-4}\,dx^0\wedge \cdots \wedge dx^{2}\wedge dz\,,
 \end{split}
\end{equation}
where $ds^2_{M_3}=-(dx^0)^2+(dx^1)^2+(dx^{2})^2$ is the metric over the three-dimensional Minkowski space $M_3$ and $\text{vol}_{4}$ the volume form on AdS$_4$ with unit radius.
Given this parametrization, we can provide an explicit form for $\hat C_5$ by expanding over 2-cycles $\{w_i\}$ as 
\begin{equation}\label{onshellC}
 \hat C_5=-\hat e^a\,z^{-3}\,\text{vol}_{M_3}\wedge w_a\qquad \text{with}\qquad \hat F_6=d\hat C_5=-\,3\,\hat e^a\,\, \text{vol}_{4}\wedge w_a \,,
\end{equation}
where $\text{vol}_{M_3}$ is the Minkowski volume form. The parameters $\hat e^a$ are constant coefficients associated to the expansion over the 2-cycles of the $F_6$ flux. Let us consider the same flux variation for all the parameters $\hat e^a$. This requirement is analogous to that one for $\tau$-variations of 2-cycles in \eqref{taudef}, given by $v^a=e^{2\tau}\hat v^a$.  We can introduce a scaling parameter $\alpha$ and a new flux configuaration $e^a$ such that
\begin{equation}\label{evariation}
 e^a=e^{\alpha}\,\hat e^a\,.
\end{equation}
This restriction is reasonable since in the large volume limit DGKT vacua are described by a single flux parameter associated to $F_4$. We can thus write the following on-shell configuaration
\begin{equation}
 \text{on-shell}:\qquad F_6=e^{\alpha}\,\hat F_6=-3\,e^a\, \text{vol}_{4}\wedge w_a \qquad \text{with} \qquad F_4=\star_{10} F_6\,.
\end{equation}
The above flux configuaration satisfies the equations of motion of massive type IIA supergravity in presence of the orientifold source.

To take flux variations, we need to promote $\alpha$ in \eqref{evariation} to a function over AdS$_4$. This assumption generates the variation of the $F_6$ flux,
\begin{equation}\label{deformedF6}
 F_6=dC_5=-e^a\left(3\,-z\,\partial_z\alpha\right) \text{vol}_{4}\wedge w_a  \,.
\end{equation}
We stress that variations \eqref{deformedF6} are defined only by derivatives along the $z$ direction. This will lead to a 4d kinetic term in the action depending only on this direction. We thus conclude that, even without imposing any restriction on the spacetime dependence of $\alpha$, our prescription singles out the AdS radial direction as special. 

The variations of the internal $F_4$ flux are defined by the Hodge dual $F_4=\star_{10}F_6$, with $F_6$ given in \eqref{deformedF6}. To obtain the explicit form of $F_4$, we first need to specify the metric variations. In this regard, it is shown in \cite{Li:2023gtt} that one has to pick a particular type of external metric variation in order to reproduce a well-defined off-shell action. In particular, the prescription for the metric must have the form \cite{Li:2023gtt}
\begin{equation}\label{AdS4sigma1}
\begin{split}
 &ds_{10}^2=e^{2\sigma} d\hat s^2_{4}+e^{2\tau}d\hat s^2_6\,,\qquad \quad d\hat s^2_4=\frac{1}{z^2}\left(ds^2_{M_3}+e^{2\sigma_1}dz^2\right) \,,
 \end{split}
\end{equation}
where we turned on a new function $\sigma_1$. 
We stress that if we take $\sigma$ and $\sigma_1$ constants, then the metric $ds^2_4=e^{2\sigma} d\hat s^2_{4}$ in \eqref{AdS4sigma1} is the AdS$_4$ metric with radius\footnote{This can be explicitly verified by reparametrizing the metric as $\bar z=e^{\sigma_1}\,z$.} $e^{2\sigma+2\sigma_1}$. So, on-shell $\sigma_1$ can be reabsorbed into $\sigma$ without loss of generality. However, when $\sigma$ and $\sigma_1$ are gauged, these variations are not equivalent to variations with $\sigma_1=0$. We will restrict to the case $\sigma_1=\sigma_1(z)$ because of the particular form of flux variations \eqref{deformedF6}, which contribute to the action only with derivatives of $\alpha$ along $z$.

We point out that the on-shell relations \eqref{onshell:dilaton} and \eqref{onshell:tau} must take into account the shift of the AdS radius by the $\sigma_1$ factor. So, for external variations of the type \eqref{AdS4sigma1}, the DGKT on-shell conditions take the form
\begin{equation}\label{onshellsigma1}
 \text{on shell:}\qquad \Phi=-(\sigma+\sigma_1)+\frac12\log\left(\frac{25}{2m^2}\right)\,,\qquad  \tau=\frac{1}{3}(\sigma+\sigma_1)+\frac16\log\left(\frac{q^2}{32}\right)
\end{equation}
Given the metric deformation \eqref{AdS4sigma1}, we can explicitly compute the Hodge dual $F_4=\star_{10}F_6$ obtaining the following result
\begin{equation}\label{deformedF4}
 F_4=\left(1-\frac{z}{3}\partial_z\alpha  \right)e_a(z)\tilde w^a\qquad \text{with}\qquad e_a\equiv12\,e^{\alpha-4\sigma-\sigma_1+6\tau}\, g_{ab}\,\hat e^b\,,
\end{equation}
where we used \eqref{dualomega} to compute the Hodge dual of $w_a$ and $g_{ab}$ is defined in \eqref{Kahlermetric}. We point out that when we evaluate \eqref{deformedF4} on-shell we obtain exactly the DGKT flux $F_4=e_a\tilde w^a$. Finally, from \eqref{onshellCondFlux} we can obtain an explicit on-shell relation expressing flux variations $\alpha$ in terms of metric and dilaton variations. Specifically, \eqref{onshellCondFlux} can be rewritten as\footnote{This condition can be obtained expressing \eqref{onshellCondFlux} in terms of ``hatted" quantities $v^a=e^{2\tau}\hat v^a$ and using \eqref{dualomega} with $\hat{\text{vol}}=1$.}
\begin{equation}\label{onshellalpha}
\text{on shell:}\qquad \qquad \alpha=3\sigma+2\tau-\Phi+\log\left(\frac{5}{3\sqrt2 m}\right)\,,
\end{equation}
where we used the on-shell relation for $\Phi$ given in \eqref{onshellsigma1}.

\subsection{Off-shell actions}

In this section we compute the off-shell actions for metric and flux variations. Since the flux variations are defined only by derivatives along the AdS radial direction $z$, we will restrict the spacetime dependence of metric variations, namely $\sigma=\sigma(z)$, $\tau=\tau(z)$ and $\Phi=\Phi(z)$. These assumptions are mandatory in our procedure otherwise  we cannot sum up the various contributions to kinetic terms and get the total metric\footnote{For a more detailed discussion on this issue we refer to Section 3 in \cite{Li:2023gtt}.}.

Let us start by summarizing our prescription for field variations
  \begin{equation}\label{variationAnsatz}
\begin{split}
&ds_{10}^2=e^{2\sigma(z)} d\hat s^2_4+e^{2\tau(z)}d\hat s^2_6\qquad \text{with}\qquad d\hat s^2_4=\frac{1}{z^2}\left(ds^2_{M_3}+e^{2\sigma_1(z)}dz^2\right)\,,\\
&F_4=\left(1-\frac{z}{3}\partial_z\alpha  \right)e_a(z)\tilde w^a\,, \qquad \qquad \Phi=\Phi(z)\,,
  \end{split}
 \end{equation}
where the ans\"atze for metric and flux variations were respectively introduced in \eqref{internalCYvariations}, \eqref{AdS4sigma1} and \eqref{deformedF4}.
We want now to compute the contributions to the 4d effective action of the above variations. Our strategy is to derive them by evaluating the type IIA action over \eqref{variationAnsatz}. In the derivation of the flux action, we will crucially ask that the on-shell conditions hold also when the variations are gauged. This requirement will allow us to reduce the total action to a single kinetic term associated to the change of the vacuum energy.

Since the above variations involve only the metric, the dilaton and the internal flux $F_4$, we can restrict our attention to the following IIA action terms:
\begin{equation}
\begin{split}\label{totalaction}
 &S_{\text{IIA}}=S_{\text{EH}}+S_{\Phi}+S_{F_4}+\cdots\,,\\
  &S_{grav}=S_{\text{EH}}+S_{\Phi}=\frac{1}{2\kappa_{10}^2}\,\int d^{10}x\sqrt{-g}\,e^{-2\Phi}\left(R+4g^{mn}\partial_m\Phi\partial_n\Phi   \right)\,,\\
  &S_{flux}=-\frac{1}{2\kappa_{10}^2}\,\int d^{10}x\sqrt{-g}\,|F_4|^2 \,.\\
  \end{split}
  \end{equation}
  We first evaluate separately $S_{grav}$ and $S_{flux}$ over the field variations \eqref{variationAnsatz}. In the next section we will combine these results together. We point out that we will cast the result in the 4d Einstein frame. As explained in Section \ref{internalvar} this procedure gives a further contribution to the metric over variations.
  
  \subsubsection{The Einstein-Hilbert and dilaton actions}
  
  Let us consider the action $S_{grav}$ evaluated over the metric variations in \eqref{variationAnsatz}. In the previous section we mentioned that metric variations \eqref{AdS4sigma1} are needed to reproduce a well-defined action. In this regard, we point out that the unique effect in the action due to $\sigma_1$ is the emergence of a term linear in $\partial_z\sigma_1$. This term will be crucial to cancel an analogous contribution coming from the flux action.
  
The linear term is extracted from the Einstein-Hilbert action. To this aim, a very useful formula is the Ricci scalar $\hat R_4$ for the metric $d\hat s^2_4=z^{-2}(ds^2_{M_3}+e^{2\sigma_1}dz^2)$. In generic $d$-dimensions one can show that \cite{Li:2023gtt}
  \begin{equation}\label{Riccisigma1}
  \hat R_d=\bar R_d-2pe^{-2\sigma_1}z\,\partial_z\sigma_1 \,,
\end{equation}
where $\bar R_d=-d(d-1)e^{-2\sigma_1}$ and $p=d-1$ is the dimension of the Minkowski foliation $M_p$. Using this formula one can verify that the inclusion of $\sigma_1$ only produces a new linear term in $\partial_z\sigma_1$ from the 4d Einstein-Hilbert action. 

The complete dimensional reduction of the Einstein-Hilbert term evaluated over $\sigma_1$-dependent variations was performed in \cite{Li:2023gtt}. In this present situation, there are also dilaton variations, but the substantial aspects of the calculation remain the same. In Section \ref{generalvar} we wrote the result of the dimensional reduction of the action $S_{grav}$ in the case $\sigma_1=0$. In that case the metric variations were described by the 4d action \eqref{genvaraction4D}. After the inclusion of $\sigma_1$, the result remains the same, except for the Ricci scalar $\hat R_4$, which is now given by formula \eqref{Riccisigma1}. Reminding that the Einstein frame metric was defined as $ds^2_4=e^{2D}ds^2_E$ with $e^{2D}=e^{2\Phi-6\tau}$, we obtain the following off-shell action
\begin{equation}
\begin{split}\label{offshellactionMetric}
 S_{4,grav}=\,\frac12\int d^4x\,\,\sqrt{- g_{E}}&\bigl(\tilde R_{E}-6e^{-2(\sigma+\sigma_1)+2D}z\,\partial_z\sigma_1+6\,( \partial \,\sigma)^2\\
 &-24\,( \partial\tau)^2+12(\partial \Phi)(\partial\tau) -2\,( \partial \Phi)^2\bigr)\,,
 \end{split}
\end{equation}
where we defined $\tilde R_{E}=e^{-2\sigma}\bar R_E$ and we used the notation $(\partial \sigma)^2=g_E^{zz}(\partial_z\sigma)^2$. In the above formula we expressed the internal volume variations in terms of the 10d dilaton $\Phi$, using formula \eqref{EinsteinFrameActioninternal}. General expressions for the metric components $K_{\sigma\sigma}$, $K_{\tau\tau}$, $K_{\Phi\Phi}$, $K_{\Phi\tau}$ have been obtained in \eqref{EHoffshell} and \eqref{internalMetric}, and these are restricted to the case $d=4$, $k=6$. We point out that in the above expression we have not written the contribution associated to the curvature of the internal space since it vanishes on-shell\footnote{We are considering vacua with smeared orientifold. In this case the internal manifold is Ricci-flat.}.

  \subsubsection{The flux action}

Let us now evaluate the flux action $S_{flux}$ introduced in \eqref{totalaction} over the variations \eqref{variationAnsatz}. We need to compute the following quantity
\begin{equation}
 S_{flux}=-\frac{1}{2\kappa_{10}^2}\,\int d^{10}x\sqrt{-g}\,|F_4|^2=-\frac{1}{2\kappa_{10}^2}\,\int d^{4}x\sqrt{-g_4}\left(1-\frac{z}{3}\partial_z\alpha  \right)^2e_ae_b\int \tilde  w^a\wedge \star_6 \,\tilde w^b,
\end{equation}
where the second intergral is performed over the internal manifold. Using the first identity in \eqref{inversemetric} and going to Planck units, we find the following 4d action
\begin{equation}\label{Sflux4d}
 S_{4,flux}=-\frac{1}{8}\,\int d^{4}x\sqrt{-g_E}e^{4\Phi-18\tau}\left(1-\frac{z}{3}\partial_z\alpha  \right)^2g^{ab}e_ae_b\,,
\end{equation}
where we introduced the Einstein frame metric $ds_4^2=e^{2D}ds^2_E=e^{2\Phi-6\tau}ds^2_E$. We point out that if we evaluate the above action on-shell, imposing $\alpha$ constant, we obtain the contribution of the $F_4$ flux to the $\ma N=1$ scalar potential. This can be immediately verified for the $T^6/\mathbb{Z}^2_3$ model where $g^{ab}e_ae_b=4 \,e_a^2v_a^2$. Imposing this relation, the action \eqref{Sflux4d} reproduces exactly the second term in the potential \eqref{T6potential}.

The action \eqref{Sflux4d} can be casted in a very simple form by imposing the on-shell condition for fluxes $e_a=-\frac{3m}{10}\kappa_{abc}v^bv^c$ given in \eqref{onshellCondFlux}. Substituting this relation into the term $g^{ab}e_ae_b$ in \eqref{Sflux4d} and using the second identity in \eqref{inversemetric} together with relations \eqref{fundamentalkappa}, we obtain the following result
\begin{equation}
 S_{4,flux}=-\frac{1}{2}\,\int d^{4}x\sqrt{-g_E}e^{4\Phi-6\tau}\left(\frac{27m^2}{25}+\frac{3m^2}{25}z^2(\partial_z\alpha)^2-\frac{18m^2}{25}z\partial_z\alpha  \right)\,.
\end{equation}
Consider the quadratic term in $\partial_z\alpha$. Using the on-shell relation $e^{2\Phi}=\frac{25}{2m^2}e^{-2(\sigma+\sigma_1)}$, given in \eqref{onshellsigma1}, we can write
\begin{equation}
 \frac{3m^2}{25}e^{4\Phi-6\tau}z^2(\partial_z\alpha)^2=\frac{3}{2}e^{-2(\sigma+\sigma_1)+2D}z^2(\partial_z\alpha)^2=\frac{3}{2}(\partial\alpha)^2\,,
 \end{equation}
where we used the notation $(\partial\alpha)^2=g_E^{zz}(\partial_z\alpha)^2$ with $g_E^{zz}=e^{-2(\sigma+\sigma_1)+2D}z^2$. If we perform a similar manipulation on the linear term in $\partial_z\alpha$, we obtain the final expression
 \begin{equation}\label{Sflux4dFinal}
 S_{4,flux}=\frac{1}{2}\,\int d^{4}x\sqrt{-g_E}\left(-\frac{27m^2}{25}e^{4\Phi-6\tau}-\frac{3}{2}(\partial\alpha)^2+9e^{-2(\sigma+\sigma_1)+2D}z\partial_z\alpha  \right)\,.
\end{equation}
We thus observe that in addition to the kinetic term, a linear term in $\partial_z\alpha$ is present in the off-shell action. In what follow we join \eqref{Sflux4dFinal} with the action associated to metric variations \eqref{offshellactionMetric}. As we will see, we will be able to cancel the two linear terms in $\partial_z\alpha$ and  $\partial_z\sigma_1$, reproducing a well-defined total metric over field variations.

\subsection{The metric over DGKT vacua}

We are now ready to combine the action for metric and flux variations, given in \eqref{offshellactionMetric} and \eqref{Sflux4dFinal}. We obtain the following 4d off-shell action
\begin{equation}
\begin{split}\label{totalactionLinearTerms}
  S_{4}=\frac{1}{2}\,\int d^{4}x\sqrt{-g_E}&\bigl(\tilde R_{E}-\frac{27m^2}{25}e^{4\Phi-6\tau}+6\,( \partial \,\sigma)^2-24\,( \partial\tau)^2+12(\partial \Phi)(\partial\tau)-2\,( \partial \Phi)^2\\
  &-\frac{3}{2}(\partial\alpha)^2-6e^{-2(\sigma+\sigma_1)+2D}z\,\partial_z\sigma_1+9e^{-2(\sigma+\sigma_1)+2D}z\partial_z\alpha  \bigr)\,.
  \end{split}
\end{equation}
As we pointed out in Section \ref{sec:fluxvariations}, the metric variations in \eqref{variationAnsatz} are underdetermined on-shell since $\sigma_1$ describes the shifts of the AdS radius. This gives us the freedom to choose $\sigma_1$. If we look now at the two linear terms in $\partial_z\alpha$ and $\partial_z\sigma_1$ in \eqref{totalactionLinearTerms}, we conclude that they cancel if we impose
\begin{equation}\label{sigma1cond}
 \sigma_1=\frac32\,\alpha\,.
\end{equation}

Let us summarize the on-shell conditions \eqref{onshellsigma1} and \eqref{onshellalpha},
\begin{equation}
 \begin{split}
 &\tau=\frac{1}{3}(\sigma+\sigma_1)+\frac16\log\left(\frac{q^2}{32}\right)\,,\qquad \Phi=-(\sigma+\sigma_1)+\frac12\log\left(\frac{25}{2m^2}\right)  \,,\\
  & \alpha=3\sigma+2\tau-\Phi+\log\left(\frac{5}{3\sqrt2 m}\right)\,.
 \end{split}
\end{equation}
Complemented with relation \eqref{sigma1cond}, the above conditions constitute a system of four equations with five parameters. We can thus solve them with respect to $\sigma$, obtaining the following result
\begin{equation}
 \begin{split}\label{finalConditions}
 &\tau=-\frac{11}{9}\sigma+\frac{1}{18}\log\left(\frac{3^6q^2}{2^5}\right)\,,\qquad \Phi=\frac{11}{3}\sigma+\log\left(\frac{5\,q^{2/3}}{2^{1/6}\,12m}  \right)\,,\\
 &\sigma_1=-\frac{14}{3}\sigma+\frac{1}{3}\log\left(\frac{2^5\,3^3}{q^2}\right)\,,\qquad \alpha=-\frac{28}{9}\sigma+\frac{2}{9}\log\left(\frac{2^5\,3^3}{q^2}\right)\,.
 \end{split}
\end{equation}
We can finally evaluate the off-shell action \eqref{totalactionLinearTerms} over the constraints \eqref{finalConditions}. The action then boils down to
\begin{equation}
\begin{split}
  S_{4}=\frac{1}{2}\,\int d^{4}x\sqrt{-g_E}&\bigg(\tilde R_{E}-\frac{27m^2}{25}e^{4\Phi-6\tau}-K_{\text{DGKT}}\,( \partial \,\sigma)^2  \bigg)\,,
  \end{split}
\end{equation}
where in the expression above we introduced the total metric $K_{\text{DGKT}}$ over the space of variations of DGKT vacua, which is given by
\begin{equation}\label{DGKTmetric}
 K_{\text{DGKT}}=\frac{3376}{27}\,.
\end{equation}
We point out that the above metric is positive, and therefore leads to a well-defined AdS distance. It is also important to stress that the above result does not represent the mere sum of the contributions of metric and flux variations. Instead, metric and flux variations combine themselves non-trivially in \eqref{finalConditions} to give rise a one-parameter family of variations with a positive metric.\footnote{The metric (\ref{DGKTmetric}) can also be written in terms of the $F_4$ flux scaling parameter $F_4 \sim n$. We can trade $\sigma$ for $n$ using the scaling $e^{2\sigma}\sim n^{3/2}$ given in \eqref{DGKTscalings}, obtaining $ K_{\text{DGKT}\,,nn}=\frac{211}{3}$.}

Finally, we can cast the above result in the standard language of the Distance Conjecture. The AdS Distance Conjecture for DGKT vacua (in Planck units) takes the form \cite{Lust:2019zwm,Junghans:2020acz}
\begin{equation}\label{DGKTdistance}
 m_{\mathrm{KK}}\sim |\Lambda|^\delta \qquad \text{with} \qquad \delta=\frac{7}{18}\,,
\end{equation}
where $m_{\mathrm{KK}}$ is the mass scale of the tower of KK states. The exponent $\delta$ is determined by the fundamental scaling behaviors of DGKT vacua in the large CY volume limit. In this limit, these take the form $\Lambda\sim n^{-9/2}$ and $m_{\text{KK}}\sim n^{-7/4}$ with $n\sim e_a$ \cite{DeWolfe:2005uu,Junghans:2020acz}. We can now compare \eqref{DGKTdistance} with the standard prescription for the mass scale of the tower of states predicted by the Distance Conjecture $m_{\text{KK}}\sim e^{-\gamma \Delta}$, where $\Delta$ is the distance within the field space. In our approach, the parameter $\gamma$ is precisely fixed by the computation of the metric over vacua variations. For DGKT vacua we have
\begin{equation}
 \gamma=\frac{7}{9\sqrt K}\,,
\end{equation}
where we traded $\sigma$ with $\Lambda$ variations using that $\Delta=-\frac12 \int \sqrt K d\log\Lambda$ (see equation (22) in \cite{Lust:2019zwm}) and we used $\delta=\frac{7}{18}$. Taking the numerical value of $ K_{\text{DGKT}}$ from \eqref{DGKTmetric}, we obtain $\gamma \simeq 0.06$. Note that this value is below the lower bound on $\gamma$ proposed for the Distance Conjecture restricted to the moduli space \cite{Grimm:2018ohb,Etheredge:2022opl}.

\section{Summary}
\label{sec:dis}

In this paper we studied the notion of a metric and a distance over families of AdS solutions in string theory. Specifically, we applied the prescription for calculating such a metric, developed in \cite{Li:2023gtt}, to the infinite one-parameter family of solutions found in DGKT \cite{DeWolfe:2005uu}. Famously, these are only smeared approximate solutions, and it remains an open question whether they exist as truly local solutions of string theory, but our procedure allows to calculate the metric in the smeared approximation. Our calculation yielded the metric (\ref{DGKTmetric}), which then implies an associated notion of distance between vacua in the family. Of particular interest to us was the sign of the metric, which turns out to be positive.

 The positivity of the metric has a number of possible implications. First, one can view it as a certain consistency test for DGKT that has been passed. Interestingly, the metric contributions coming from variations of the spacetime metric only, yield a negative result. But upon including the contributions from the dilaton and flux variations, the final result is positive. 
 
 Another implication is that the result seems to make more robust the idea that the procedure for calculating metrics over families of solutions yields sensible results. A particularly nice feature is that it turns out that the metric is independent of any of the various flux parameters in the solution. In this sense, it behaves like other metrics we see in string theory over scalar field spaces. Encouraged by these nice properties and results, we proposed a Metric Positivity Conjecture which states that metrics over families of solutions in quantum gravity always have to be positive (or at least yield real proper distances in the space of solutions). Of course, this conjecture assumes a procedure for calculating the metrics, and we only  have a sharp specific procedure for doing so for simple families of AdS vacua in string theory. Nonetheless, the underlying ideas behind calculating the metric, of taking an off-shell quadratic variation of the action and then restricting this using the on-shell relations between the parameters, can be considered general. Even if how to implement this precisely in more complicated solutions remains an open question. 
 
 For other relatively simple solutions in string theory, the conjecture and calculating procedure are sharp. We therefore aim to perform these calculations in future work. For example, there exist two-parameter type IIA vacua of the type AdS$_4\times Y_6$. This analysis should hopefully also yield information on how to define the metric and paths over multi-parameter spaces.

\vskip 20pt
	\noindent {\bf Acknowledgements:} We would like to thank Joan Quirant for useful comments and discussions. EP and NP are supported by the Israel Science Foundation (grant No. 741/20) and by the German Research Foundation through a German-Israeli Project Cooperation (DIP) grant ``Holography and the Swampland". EP is supported by the Israel planning and budgeting committee grant for supporting theoretical high energy physics.
\newpage
\appendix

\section{Type IIA orientifolds: the moduli space}\label{appendixModuliSpace}

In this appendix we review the main properties of moduli spaces from type IIA orientifold compactifications. As this is a very standard topic in string compactification, we will mainly follow the analysis of the seminal papers \cite{DeWolfe:2005uu, Grimm:2004ua}. For aspects of $\ma N=2$ supergravity from type IIA compactifications, we will refer to \cite{Louis_2002}.

It is well known that moduli fields describe the variations of the geometry of the Calabi-Yau. To study these variations, the standard procedure in string compactification consists in expanding the internal fluxes over harmonic forms, which are in one-to-one relation with internal cycles. 

The $\ma N=1$ compactifications are usually performed in two steps. First one extracts the $\ma N=2$ supergravity arising from reducing type IIA supergravity over Calabi-Yau threefolds. Then one excludes those fields that do not preserve the orientifold symmetry. This procedure leads to the $\ma N=1$ moduli space in which the size of the two sectors of the $\ma N=2$ moduli space, K\"ahler and complex structure, has been reduced.

\subsection{K\"ahler moduli space}\label{KahlerModuli}

The  K\"ahler moduli space describes variations of the complexified K\"ahler form $J_c=B_2+i J$. The deformations of the K\"ahler form $J$ are typically associated to the volume of 2-cycles, while variations of the $B_2$ field are described by axionic modes. The form $J_c$ can be expanded in the basis of harmonic (1,1)-forms. This expansion is described by $h^{1,1}$ complex moduli organized into $\ma N=2$ vector multiplets.

The form $J_c$ is odd under the orientifold. This implies that moduli fields surviving under the projection must be associated to odd (1,1)-forms. If we call $h^{1,1}_-$ and $h^{1,1}_+=h^{1,1}-h^{1,1}_-$, the dimensions of subspaces of odd and even harmonic $(1,1)$-forms, the expansion of $J_c$ is given by
\begin{equation}
 J_c=t^aw_a \qquad \text{with} \qquad t^a=b^a+iv^a\qquad \text{and}\qquad a=1,\dots,h^{1,1}_-\,,
\end{equation}
where $\{w_a\}$ is a basis of odd harmonic (1,1)-forms. The K\"ahler potential $K^K$ can be obtained by restricting the $\ma N=2$ potential to the odd moduli,
\begin{equation}\label{KahlerKK}
 e^{-K^K}=8\,\text{vol}=\frac43\,\kappa_{abc}v^av^bv^c\,,
\end{equation}
where we used the expansion $J=v^aw_a$ and we introduced the volume modulus
\begin{equation}\label{CYvolume}
 \text{vol}=\frac16\,\int J\wedge J\wedge J \qquad \text{and}\qquad \kappa_{abc}=\int w_a\wedge w_b\wedge w_c\,.
\end{equation}
The numeric coefficient $\kappa_{abc}$ is called the intersection number of the Calabi-Yau.
We can then introduce the following quantities
  \begin{equation}\label{fundamentalkappa}
   \kappa_{ab}=\kappa_{abc}v^c\qquad \text{and}\qquad  \kappa_{a}=\kappa_{ab}v^b\qquad \text{with}\qquad \kappa_{a}v^a=6\,\text{vol}\,. 
  \end{equation}
Using these quantities we can write the K\"ahler metric $g_{ab}=\partial_a\bar{\partial}_bK^K$ as it follows
\begin{equation}\label{Kahlermetric}
 g_{ab}=\frac{1}{4\,\text{vol}}\,\int \,w_a\wedge \star_6\,w_b=-\frac{1}{4\text{vol}}\,\left(\kappa_{ab}-\frac14\,\frac{\kappa_a\kappa_b}{\text{vol}} \right)\,.
\end{equation}
In a Calabi-Yau threefold odd 2-cycles are dual to even 4-cycles, implying that $h^{1,1}_-=h^{2,2}_+$. It follows that fluxes along 4-cycles can be expanded over bases of even (2,2)-forms. We can introduce such a basis $\{\tilde w^a\}$ with $\int w_a \wedge \tilde w^b=\delta_a^b$. More explicitly, bases of 2- and 4-cycles are related each other as it follows,
\begin{equation}\label{dualomega}
 \star_6w_a=4\,\text{vol}\,g_{ab}\tilde w^b\qquad \text{and} \qquad \star_6\tilde w^a=\frac{1}{4\text{vol}}\,g^{ab} w_b\,.
\end{equation}
We point out that in the second relation we introduced the inverse metric
\begin{equation}\label{inversemetric}
 g^{ab}=4\,\text{vol}\int\,\tilde w^a\wedge \star_6\,\tilde w^b=-4\text{vol}\left(\kappa^{ab}-\frac{v^av^b}{2\text{vol}}\right)\,,
\end{equation}
where in the last equality $\kappa^{ab}$ is defined as the quantity such that $\kappa^{ab}\kappa_{bc}=\delta^a_c$.

\subsection{Complex structure moduli space}\label{hyperModuli}

Let's consider the moduli space sector associated to variations of 3-cycles. In $\ma N=2$ compactifications this is described by the complex structure moduli space and it is associated to deformations of the holomorphic 3-form
\begin{equation}\label{omega}
 \Omega=Z^{\hat{K}} \alpha_{\hat{K}} -\ma F_{\hat{L}}\,\beta^{\hat{L}}\,,
\end{equation}
 with $\hat{K},\hat{L}=0,\dots,h^{2,1}$. The forms $\{\alpha_{\hat{K}}, \beta^{\hat{L}}\}$ constitute a basis of harmonic 3-forms, namely $\int \alpha_{\hat K}\wedge \beta^{\hat L}=\delta^{\hat L}_{\hat K}$. The periods $Z^{\hat{K}}(z)$ are the homogenous coordinates over the complex structure\footnote{The periods $\ma F_{\hat{L}}$ are dependent to $Z^{\hat K}$ as they can be written explicitly as the derivatives of a prepotential $\ma F(Z)$, namely $\ma F_{\hat{L}}=\partial_{\hat{L}} \,\ma F (Z)$.} and define the physical moduli as $z^K=Z^K/Z^0$, with $ K=1,\dots,h^{2,1}$. Each of the complex moduli $z^K$ can be complemented by two axions to build up a $\ma N=2$ hypermultiplet. These axions, usually called  $\xi^{\hat{K}}$ and $\tilde \xi_{\hat{K}}$, belong to the expansion of the internal 3-form gauge potential $C_3$,
\begin{equation}\label{C3}
 C_3=\xi^{\hat{K}} \alpha_{\hat{K}}-\tilde \xi_{\hat{L}}\beta^{\hat{L}}\,.
\end{equation}
Overall, the complex structure moduli space is composed by $h^{2,1}$ hypermultiplets $(z^K,\xi^K,\tilde \xi_K)$, plus the universal hypermultiplet. The latter is composed by two axionic modes, $\xi^0$, $\tilde \xi_0$, the 10d dilaton $\Phi$ and the dual mode of the spacetime component of the $B_2$ field. One has then $2(h^{2,1}+1)$ complex moduli.

Let's consider now the orientifold projection. We point out that 3-forms decompose independently along even and odd bases, namely $h^{3}_+=h^{3}_-=h^{2,1}+1$. Hence, before the orientifold, one can split a basis of harmonic 3-forms by choosing all the $\alpha_ {\hat K}$ even and the $\beta^{\hat L}$ odd. Such a basis can be always obtained through a symplectic rotation allowed by $\ma N=2$ supersymmetry. The orientifold breaks the symplectic invariance, fixing a particular frame for the 3-form basis. Concretely, one can write such frame of even and odd bases as $\{\alpha_k, \beta^\lambda \}$ and $\{\alpha_\lambda, \beta^k \}$ with $k=0,1,\dots,h$ and $\lambda=h,\dots, h^{2,1}$. The parameter $h$ gives the number of even $\alpha$s in the expansion and it is fixed by the specific frame chosen.

Given the above splitting of 3-forms, one can consider the orientifold action on $\Omega$ obtaining that $\Im\, Z^k=\Re \,\ma F_k=\Re \,Z_\lambda=\Im \ma \,F_\lambda=0$. After imposing these relations, only one real component of each modulus $z^K$ survives \footnote{Two of these relations are redundant.}. The same procedure applied to the potential $C_3$ projects out one of the two axions in each hypermultiplet. It follows that the expansions \eqref{omega} and \eqref{C3} boil down to
\begin{equation}
 \Omega=(\Re Z^{k}\alpha_{k}+i\Im Z^\lambda\alpha_{\lambda})  - (\Re \ma F_{\lambda}\beta^{\lambda}+i\Im \ma F_k\beta^{k})\qquad \text{and} \qquad C_3=\xi^k\alpha_k-\tilde \xi_\lambda \beta^\lambda.
\end{equation}
We then obtain that each hypermultiplet is cut in half and only $h^{2,1}+1$ complex moduli survive. Specifically, as far as the universal hypermultiplet is regarded, only the 10d dilaton $\Phi$ and one axion among $\xi^0$ and $\tilde \xi_0$ is retained, we will call it $\xi$. 

The surviving moduli define a complex structure over the subspace preserved by the orientifold. Usually, these moduli are organized in the following form
\begin{equation}\label{complexStructureNT}
 N^k=\frac{1}{2}\,\xi_k+i \Re \left(e^{-D+K^{\text{cs}}/2} \,Z^k\right)\,,\qquad \qquad T_\lambda=i\,\tilde \xi_\lambda-2 \Re \left(e^{-D+K^{\text{cs}}/2}\,\ma F_\lambda \right)\,,
\end{equation}
where $e^{-K^{\text{cs}}}=2\left(\Im Z^\lambda \Re \ma F_\lambda-\Re Z^k\,\Im \ma F_k  \right)$ is the K\"ahler potential of the complex structure restricted to the surviving moduli and $e^{2D}=\frac{e^{2\Phi}}{\text{vol}}$ is the 4d dilaton. As we mentioned above, these expressions depend on the explicit basis of 3-forms chosen. For example, if one takes $h=h^{2,1}$, the unique non-zero moduli are the $N_k$. Finally, one can show that the K\"ahler potential for the surviving moduli is given by \cite{Grimm:2004ua,DeWolfe:2005uu}
\begin{equation}\label{KahlerKQ}
 K^Q=4D\,.
\end{equation}
We point out that, despite the orientifold breaks the symplectic covariance, giving rise to basis-dependent expressions for \eqref{complexStructureNT}, the above formula of the K\"ahler potential is basis-independent.

	\bibliographystyle{jhep}
	\bibliography{metricdistance}
\end{document}